\documentclass[preprint,showpacs,preprintnumbers,amsmath,amssymb,aps,color,nofootinbib]{revtex4-1}
\usepackage{braket,color,comment,enumerate,epsfig,graphicx,here,hyperref,mathrsfs,natbib,slashed,subfigure}
\def\nn{\nonumber}
\def\beq{\begin{equation}}
\def\eeq{\end{equation}}
\def\bea{\begin{eqnarray}}
\def\eea{\end{eqnarray}}

\begin{document}
%==================================
\preprint{}
\title{$D^0-\bar{D}^0$ mixing in the Dyson-Schwinger approach}
\author{Xiaotong Xie}
\email[E-mail: ]{jiext23@mails.jlu.edu.cn}
\author{Hiroyuki Umeeda}
\email[E-mail: ]{umeeda@jlu.edu.cn}
\author{Jinglong Zhu}
\email[E-mail: ]{zhujl23@mails.jlu.edu.cn}
\affiliation{Center for Theoretical Physics and College of Physics, Jilin University, Changchun, 130012, China}
\date{\today}
\begin{abstract}
In view of difficulty to reproduce observables in the $D^0-\bar{D}^0$ mixing via the operator product expansion, we discuss the Dyson-Schwinger approach to this process. Formulated by the parametrization of quark propagators, SU(3) breaking relevant to charm mixing is evaluated in such a way that properly takes account of dynamical chiral symmetry breaking. The $\bar{D}^0\to D^0$ transition is discussed in the vacuum-insertion approximation with locality of the light valence-quark field, represented by the decay constant of $D^0$ meson as well as relevant momentum integrals. It is found that dimensionless mass-difference observable in this approach leads to $|x|=(1.3-2.9)\times 10^{-3}$, the order of magnitude comparable to the HFLAV data, and thereby offering a certain improvement as a theoretical framework.
\end{abstract}
\maketitle
%============
\section{Introduction}
%============
Owing to the notable development of heavy quark physics in the 1990s, strong interaction of the $b$ quark system can be systematically handled, which enables us to analyze a variety of observables to date. Armed with those formulations, precise determinations of the Cabibbo-Kobayashi-Maskawa (CKM) matrix elements \cite{Cabibbo:1963yz}, as well as the search for physics beyond the standard model, have been implemented in recent flavor-factory experiments.
In contrast to the mature status in $B$-meson system, however, there still exist difficulties in regards to the charm sector. In particular, dominance of long-distance effects for the $D^0-\bar{D}^0$ mixing has been remarked \cite{Donoghue:1985hh}, and therefore requiring dedicated analysis incorporating nonperturbative dynamics.
\par
The conventional theoretical methods for the $D^0-\bar{D}^0$ mixing are classified as two types: The exclusive and inclusive approaches. The former offers parametrization of decay amplitudes, with specific methods given by topological diagram approach \cite{Cheng:2010rv, Cheng:2024hdo}, factorization-assisted topological approach \cite{Jiang:2017zwr} or flavor symmetry approach \cite{Gronau:2012kq} and yields the data-driven formalism to predicts the observables in the $D^0-\bar{D}^0$ mixing. This methodology takes finite sum over hadronic intermediate states of $\bar{D^0}\to \textrm{hadrons}\to D^0$, reproducing approximately a half of the experimental data relevant to width difference \cite{Cheng:2010rv, Jiang:2017zwr, Cheng:2024hdo}.
\par
As for the inclusive approach\footnote{See earlier discussions based on box diagrams in Refs.~\cite{Hagelin:1981zk, Cheng:1982hq, Buras:1984pq, Datta:1984jx} and the ones in heavy quark effective theory (HQET) in Refs.~\cite{Georgi:1992as, Ohl:1992sr}. The dispersive approach based on the inclusive analysis as input is found in Refs.~\cite{Li:2020xrz, Li:2022jxc}. Violation of quark-hadron duality is discussed in Refs.~\cite{Bigi:2000wn, Jubb:2016mvq, Umeeda:2021llf} in a generic context, the simplified model and the 't Hooft model, respectively.} \cite{Golowich:2005pt, Bobrowski:2010xg} (a recent work is found in Ref.~\cite{Melic:2024oqj}), the primary formulation is based on the operator product expansion (OPE) \cite{Wilson:1969zs}.  In a sharp contrast to the mentioned status in the exclusive approach, the inclusive one gives predictions that are much smaller than the experimental data: If we take a dimensionless mass-difference observable as an example, the comparison between the OPE value at the next-to-leading order (NLO) in Ref.~\cite{Golowich:2005pt} and the experimental result from the heavy flavor averaging group (HFLAV) \cite{HeavyFlavorAveragingGroupHFLAV:2024ctg} reads,
\bea
x_{\rm Th.}=6\times 10^{-7},\qquad
x_{\rm Exp.}=(4.07\pm 0.44)\times 10^{-3},\label{Eq:1}
\eea
where the definition of $x$ is properly introduced later in Sec.~\ref{Sec:IIA}. As shown in Eq.~(\ref{Eq:1}), there exists a gap between the two results in four orders of magnitude.
\par
The difficulty in reproducing the data for the inclusive approach is attributed to the extreme suppression that originates from the Glashow-Iliopoulos-Maiani (GIM) mechanism \cite{Glashow:1970gm}. As a consequence of this, in the limit where the product of the CKM matrix elements $(V_{cb}V_{ub}^*)$ vanishes, the observables in the $D^0-\bar{D}^0$ mixing are proportional to SU(3) breaking as was remarked in Ref.~\cite{Kingsley:1975fe}. In order to capture the relevant size of the observables, proper evaluation of {\it hadronic} (or nonperturbative) SU(3) breaking plays a crucial role. See Refs.~\cite{Bigi:2000wn, Bobrowski:2010xg, Falk:2001hx, Falk:2004wg} for further discussions.
\par
For the $c\to u\gamma$ decay, the difficulty similar to the $D^0-\bar{D}^0$ mixing is also found: In Ref.~\cite{Greub:1996wn}, branching ratio of inclusive radiative decay, $\Delta \textrm{Br}[D^0\xrightarrow{\Delta S=0} X\gamma]_{c\to u\gamma}$, is obtained as $2.5\times 10^{-8}$, much smaller than the particle data group (PDG) value of one exclusive channel, $\textrm{Br}[D^0\to \rho^0\gamma]=(1.82\pm 0.32)\times 10^{-5}$ \cite{ParticleDataGroup:2024cfk}.
For this, robust enhancement of the result is confirmed \cite{Greub:1996wn}, if one replaces the current-quark masses for $s$ and $d$ quarks by ones of the constituent-quark masses. Even though the process is different, a similar enhancement in a larger magnitude is expected for charm mixing. Furthermore, for the $D^0-\bar{D}^0$ mixing in the OPE-based approaches, the condensate contributions \cite{Bigi:2000wn}, as well as nonlocal chiral condensates \cite{Melic:2024oqj}, are discussed as power-suppressed terms that can avoid severe GIM suppression, and thereby leading to large corrections. Those previous works generically imply the importance of chiral symmetry breaking in charm mixing.
\par
In this work, we evaluate an observable in the $D^0-\bar{D}^0$ mixing based on the Dyson-Schwinger equation (DSE) approach \cite{Roberts:1994dr,Roberts:2021nhw}. Formulated by the propagators for intermediate $d$ and $s$ quarks, the dimensionless mass-difference parameter is analyzed, with SU(3) breaking properly incorporated. To this end, the parametrization \cite{Burden:1991gd, Burden:1995ve, Kalinovsky:1996ii, Ivanov:1997yg, Ivanov:1998ms} (see Refs.~\cite{El-Bennich:2011tme,El-Bennich:2016bno} for recent works) accommodating confinement and dynamical chiral symmetry breaking (DCSB) as well as asymptotic freedom is adopted. This framework enables us to estimate the size of observables with nonperturbative SU(3) breaking via the relevant momentum integrals of the quark propagators.
\par
This paper is organized as follows: In Sec.~\ref{Sec:IIA}, theoretical formulas for the $D^0-\bar{D^0}$ mixing, as well as definitions of mixing parameters, are introduced. In Sec.~\ref{Sec:IIB}, the OPE approach to the $D^0-\bar{D^0}$ mixing is briefly recapitulated. In Sec.~\ref{Sec:IIC}, SU(3) breaking that originates from the GIM mechanism is discussed with some replacements of quark masses. In Sec.~\ref{Sec:III}, the mixing parameter is estimated in the DSE approach, giving numerical results in this work. The concluding remarks are addressed in Sec.~\ref{Sec:IV}.
%============
\section{$D^0-\bar{D}^0$ mixing}\label{Sec:II}
%============
\subsection{Mixing parameters}\label{Sec:IIA}
%============
Here formulas for charm mixing are given in the $\mathit{CPT}$-conserving limit \cite{HeavyFlavorAveragingGroupHFLAV:2024ctg}. Taking the phase convention of $\mathit{CP}$ transform as $\textrm{CP}\ket{D^0}=-\ket{\bar{D}^0}$, we introduce mass eigenstates,
\bea
\ket{D_{1, 2}}&=&p\ket{D^0}\mp q\ket{\bar{D}^0},
\eea
where $|p|^2+|q|^2=1$. In the $\mathit{CP}$-conserving case of $p=q$, $D_1~(D_2)$ is reduced to a $\mathit{CP}$-even ($\mathit{CP}$-odd) state. The matrix element for the $\bar{D}^0\to D^0$ transition is divided into the dispersive and absorptive parts,
\bea
M_{12}-\frac{i}{2}\Gamma_{12}&=&
\frac{\braket{D^0|\mathcal{H}_{\rm eff}|\bar{D}^0}}{2M_{D^0}},\qquad
\mathcal{H}_{\rm eff}=\mathcal{H}_{\rm disp}-\frac{i}{2}\mathcal{H}_{\rm abs},
\eea
or equivalently,
\bea
M_{12}&=&\frac{\braket{D^0|\mathcal{H}_{\rm disp}|\bar{D}^0}}{2M_{D^0}},\qquad
\Gamma_{12}=\frac{\braket{D^0|\mathcal{H}_{\rm abs}|\bar{D}^0}}{2M_{D^0}}.
\eea
In our convention, $D^0$ and $\bar{D}^0$ states are relativistically normalized,
\bea
\braket{D^0(\mathbf{p})|D^0(\mathbf{p}^\prime)}=
\braket{\bar{D}^0(\mathbf{p})|\bar{D}^0(\mathbf{p}^\prime)}=2E_{\mathbf{p}}(2\pi)^3\delta^{(3)}(\mathbf{p}-\mathbf{p}^\prime).
\eea
The observables are defined by,
\bea
&x=\displaystyle\frac{\Delta m}{\Gamma},\qquad  y=\displaystyle\frac{\Delta \Gamma}{2\Gamma},&\\
&\Delta m=m_1-m_2,\qquad
\Delta \Gamma=\Gamma_1-\Gamma_2,\qquad
\Gamma=\displaystyle\frac{\Gamma_1+\Gamma_2}{2}.&
\eea
In what follows, we take the $\mathit{CP}$-conserving limit, leading to,
\bea
&M_{12}=\textrm{Re}
\displaystyle\frac{\braket{D^0|\mathcal{H}_{\rm eff}|\bar{D}^0}}{2M_{D^0}},\qquad
\Gamma_{12}=-2\textrm{Im}\frac{\braket{D^0|\mathcal{H}_{\rm eff}|\bar{D}^0}}{2M_{D^0}}, &\\
&\Delta m=2M_{12},\qquad\Delta \Gamma=2\Gamma_{12}.&
%\\
%&\Delta m=\textcolor{red}{\pm ?}2M_{12},\qquad
%\Delta \Gamma=\textcolor{red}{\pm ?}2\Gamma_{12}.&
\eea
%\textcolor{red}{The signs need to be checked.}\\
The effective Hamiltonian for the $\Delta C=1$ transition is defined by,
\bea
\mathcal{H}_{\rm eff}&=&\frac{(-i)^2}{2!}i\int \textrm{T}[\mathcal{H}_{\Delta C=1}(x)\mathcal{H}_{\Delta C=1}(0)]d^4x.\label{Eq:effHdef}
\eea
In the case where CKM-suppressed contributions of $b$ quark are neglected, the Hamiltonian reads,
\bea
\mathcal{H}_{\Delta C=1}&=&\displaystyle\sum_{i=d}^s\displaystyle\sum_{j=d}^s\mathcal{H}_{ij},\\
\mathcal{H}_{ij}&=&\frac{4G_F}{\sqrt{2}}V_{ci}V_{uj}^*(C_1O_1+C_2O_2),\\
O_1&=&(\bar{c}^\alpha \gamma^\mu P_Lq^\beta_i)(\bar{q}^\beta_j \gamma_\mu P_Lu^\alpha),\label{Eq:O1ope}\\
O_2&=&(\bar{c}^\alpha \gamma^\mu P_Lq^\alpha_i)(\bar{q}^\beta_j \gamma_\mu P_Lu^\beta).\label{Eq:O2ope}
\eea
The sum over color indices denoted as $\alpha$ and $\beta$ are understood in Eqs.~(\ref{Eq:O1ope}), (\ref{Eq:O2ope}). In what follows, the dispersive part, which is of our current interest, is analyzed.
%=============
\subsection{OPE approach}\label{Sec:IIB}
%=============
We recapitulate the conventional inclusive approach \cite{Hagelin:1981zk, Cheng:1982hq, Buras:1984pq, Datta:1984jx, Golowich:2005pt, Bobrowski:2010xg} at the leading order (LO). One can implement the replacement,
\bea
\mathcal{H}_{\Delta C=1}(x)
\mathcal{H}_{\Delta C=1}(0)\to
\displaystyle\sum_{i, j}^{d, s}\mathcal{H}_{ij}(x)
\mathcal{H}_{ji}(0).
\eea
Furthermore, with $\lambda_i=V_{ci}V_{ui}^*$, the unitarity of the CKM matrix leads to,
\bea
\lambda_d+\lambda_s+\lambda_b=0.
\eea
In the limit of $\lambda_b\simeq0$, one can eliminate $\lambda_d$ and write the result only in terms of $\lambda_s$. This procedure renders Eq.~(\ref{Eq:effHdef}) recast into the form (see, {\it e.g.}, Ref.~\cite{Golowich:2005pt}),
\bea
\mathcal{H}_{\rm disp}&\simeq&-\frac{G_F^2m_c^2}{8\pi^2}\lambda_s^2
[C_2^2Q_1+2(C_2^2-2C_1C_2-N_cC_1^2)Q_2]\left(\frac{m_s^2-m_d^2}{m_c^2}\right)^2,\label{Eq:Hdisp}\\
Q_1&=&
[\bar{c}^\alpha\gamma^\mu (1-\gamma^5) u^\alpha]
[\bar{c}^\beta\gamma_\mu (1-\gamma^5) u^\beta],\label{Eq:Q1}\\
Q_2&=&
[\bar{c}^\alpha(1-\gamma^5) u^\alpha]
[\bar{c}^\beta (1-\gamma^5) u^\beta],\label{Eq:Q2}
\eea
where $Q_i~(i=1, 2)$ is the $\Delta C=2$ operator. The contribution of $Q_2$ in Eq.~(\ref{Eq:Hdisp}) is obtained by applying the equation of motion (see e.g., Ref.~\cite{Umeeda:2021llf}).
The matrix elements for Eqs.~(\ref{Eq:Q1}), (\ref{Eq:Q2}) are defined with bag parameters,
\bea
\braket{D^0|Q_1|\bar{D}^0}&=&\frac{8}{3}M_{D^0}^2f_{D^0}^2 B_1,\\
\braket{D^0|Q_2|\bar{D}^0}&=&-\frac{5}{3}
\left(\frac{M_{D^0}}{m_c+m_u}\right)^2
M_{D^0}^2f_{D^0}^2 B_2,
\eea
leading to ($N_c=3$),
\bea
M_{12}&\simeq&
-\frac{G_F^2f_{D^0}^2M_{D^0}m_c^2}{6\pi^2}\lambda_s^2
\left[C_2^2B_1-\frac{5}{4}(C_2^2-2C_1C_2-N_cC_1^2)
\left(\frac{M_{D^0}}{m_c+m_u}\right)^2B_2\right]
\left(\frac{m_s^2-m_d^2}{m_c^2}\right)^2.\qquad\label{Eq:M12SU3}
\eea
It is evident that $M_{12}$ vanishes in the SU(3) limit, $m_s=m_d$, as a consequence of the GIM mechanism with $\lambda_b=0$.
%========
\subsection{GIM suppression factor}\label{Sec:IIC}
%========
One can estimate the size of the last factor in Eq.~(\ref{Eq:M12SU3}): If one fixes masses for $s$, $d$ and $c$ to the $\overline{\rm MS}$ masses at the scale of charm quark mass, where renormalization group evolution (RGE) for $m_s$ and $m_d$ are evaluated by RunDec \cite{Chetyrkin:2000yt} with input parameters of $m_c(m_c)$, $m_d(2~\textrm{GeV})$ and $m_s(2~\textrm{GeV})$ in Ref.~\cite{ParticleDataGroup:2024cfk}, we obtain,
\bea
\left(\frac{m_s^2-m_d^2}{m_c^2}\right)^2=5.2\times 10^{-5},\label{Eq:SU3b1}
\eea
giving the extreme suppression in Eq.~(\ref{Eq:M12SU3}).
\par
Here we consider the replacement of the current-quark masses by the constituent-quark masses similarly to Ref.~\cite{Greub:1996wn} for $c\to u\gamma$ decays. Specifically, the constituent-quark mass in the relativistic constituent quark model (RCQM) \cite{Schlumpf:1992vq} and the Euclidean constituent-quark mass, obtained as a solution of $M^2(p^2_E)=p^2_E$ with $M(p^2_E)$ determined in the DSE approach \cite{Ivanov:1998ms}, are adopted for illustrating the size of the factor. If we take $m_s^{\rm con}=0.49~\textrm{GeV}~\textrm{and}~m_d^{\rm con}=0.36~ \textrm{GeV}$~\cite{Ivanov:1998ms} as an example, the factor corresponding to Eq.~(\ref{Eq:SU3b1}) is,
\bea
\left[\frac{(m_s^{\rm con})^2-(m_d^{\rm con})^2}{m_c^2}\right]^2=
4.6\times 10^{-3},\label{Eq:SU3bstar}
\eea
which is two orders of magnitude larger than Eq.~(\ref{Eq:SU3b1}). The resulting observable is also enhanced by the mentioned factor.
\par
In order to obtain the numerical result, we adopt $M_{D^0}=1.86484~\textrm{GeV}$ in PDG \cite{ParticleDataGroup:2024cfk} and $f_{D^0}=0.2120~\textrm{GeV}$ in lattice QCD \cite{Carrasco:2014poa} (see also \cite{FlavourLatticeAveragingGroupFLAG:2024oxs}).
As to the CKM matrix elements in the $\mathit{CP}$-conserving limit, we adopt the absolute values in PDG, $|V_{cs}|=0.975$ and $|V_{us}|=0.22431$ \cite{ParticleDataGroup:2024cfk} for $\lambda_s^2$ in Eq.~(\ref{Eq:M12SU3}). For the bag parameters, they are computed via lattice QCD in Refs.~\cite{Gupta:1996yt, Carrasco:2014uya, Carrasco:2015pra, Bazavov:2017weg}, with central values of Ref.~\cite{Carrasco:2015pra} 
in Eqs.~(\ref{Eq:Bag1}, \ref{Eq:Bag2}, \ref{Eq:Bag3}) adopted for our analysis. The corrections from RGE to the bag parameters and the Wilson coefficients are given in Appendix~\ref{App:A}. The $u$ quark mass in Eq.~(\ref{Eq:M12SU3}) is fixed to the $\overline{\rm MS}$ mass at the scale of $m_c$ via the RunDec \cite{Chetyrkin:2000yt}.
\par
In Table~\ref{Tab:I}, values of $|x|$ with different masses for $s$ and $d$ quarks are compared. As can be seen in Table~\ref{Tab:I}, the cases with constituent-quark masses give larger $|x|$ than that with the $\overline{\rm MS}$ masses. The resulting enhancement is highly sensitive to the values of $m_s^{\rm con}$ and $m_d^{\rm con}$.
\begin{table}[t]
\centering
\caption{The dimensionless mass-difference observable $(|x|)$ with different values of quark-mass parameters: (a) $\overline{\rm MS}$ mass. (b), (c), and (f) RCQM with the harmonic oscillator wave function. (d) constituent-quark mass adopted in $c\to u\gamma$ study. (e) and (g) the Euclidean constituent-quark mass in the DSE approach. For the constituent-quark masses, the isospin limit, $m_u^{\rm con}=m_d^{\rm con}$, is adopted.}
\label{Tab:I}
\begin{tabular}{cccccc}
\hline\hline
&  $|x|$&$m_s^{(\textrm{con})}/\textrm{GeV}$ & $m_d^{(\textrm{con})}/\textrm{GeV}$ & Method/Model& Reference\\\hline
(a) &$2.5\times 10^{-6}$  &$0.108$ & 0.0055 & $\overline{\rm MS}$ mass & \cite{ParticleDataGroup:2024cfk, Chetyrkin:2000yt}\\\hline
(b) &$1.1\times 10^{-4}$&$0.38$ & 0.26 & RCQM &\cite{Schlumpf:1992vq}\\\hline
(c) &$1.5\times 10^{-4}$&$0.40$ & 0.267 & RCQM &\cite{Schlumpf:1992vq}\\\hline
(d) &$2.3\times 10^{-4}$&$0.45$ & 0.30 & Constituent-quark mass&\cite{Greub:1996wn} \\\hline
(e) &$2.3\times 10^{-4}$&$0.49$ & 0.36 & DSE &\cite{Ivanov:1998ms}\\\hline
(f) &$6.9\times 10^{-4}$&$0.55$ & 0.33 & RCQM &\cite{Schlumpf:1992vq}\\\hline
(g) &$5.8\times 10^{-4}$&$0.70$ & 0.56 & DSE & \cite{Ivanov:1998ms}\\\hline\hline
\end{tabular}
\end{table}
\par
It should be noted, however, that momentum dependence of the quark propagators is not fully incorporated in the simple replacement for the quark masses in Eq.~(\ref{Eq:SU3bstar}) and also in the numerical results of Table~\ref{Fig:1} (b)-(g). In the next section, the integral over the relevant momenta is carried out using the DSE approach.
%============
\section{DSE approach}\label{Sec:III}
%============
\subsection{Quark propagator}
%============
The quark propagator in the Minkowski notation is denoted as,
\bea
S_i(x, y)=\int\frac{d^4p}{(2\pi)^4}i[\slashed{p}\sigma_{V}^i(-p^2)+\sigma_{S}^i(-p^2)]e^{-ip\cdot (x-y)}.\quad (i=d, s)\label{Eq:PropS}
\eea
In Eq.~(\ref{Eq:PropS}), $\sigma_{V}^i(-p^2)$ and $\sigma_{S}^i(-p^2)$ represent the vector and the scalar parts, respectively, with the sign of the argument taken as negative for the later convenience of the Euclidean notation. In Refs.~\cite{Burden:1991gd, Burden:1995ve,Kalinovsky:1996ii, Ivanov:1997yg, Ivanov:1998ms}, the parametrization of the $\sigma_{V,\: S}$ has been discussed. One can incorporate DCSB, which is not explicitly seen for the OPE at the leading power approximation, in this method. Color confinement is also accommodated in the sense that the parametrization does not possess the Lehmann representation \cite{Burden:1991gd}. The explicit forms for $\sigma_{V, \: S}$ are given by \cite{Burden:1991gd,Burden:1995ve,Kalinovsky:1996ii, Ivanov:1997yg, Ivanov:1998ms},
\bea
\sigma_V^{i}(p_E^2)&=&\frac{\bar{\sigma}_V^i(x)}{2D},\qquad
\sigma_S^{i}(p_E^2)=\frac{\bar{\sigma}_S^i(x)}{\sqrt{2D}},\label{Eq:propV}\\
\bar{\sigma}_V^i(x)&=&\frac{2(x+\bar{m}_i^2)-1+e^{-2(x+\bar{m}_i^2)}}{2(x+\bar{m}_i^2)^2},\label{Eq:propV2}\\
\bar{\sigma}_S^i(x)&=&2\bar{m}_i \mathcal{F}(2(x+\bar{m}_i^2))+
\mathcal{F}(b_1^ix)\mathcal{F}(b_3^ix)[b_0^i +b_2^i \mathcal{F}(\epsilon x)],\label{Eq:propS}
\eea
where $p_E^2=-p^2$, $x=p_E^2/(2D), \bar{m}_i=\tilde{m}_i/\sqrt{2D}$, and $\mathcal{F}(z)=(1-e^{-z})/z$ while $b_n^{i}~(n=0, 1, 2, 3), D$ and $\epsilon$ are parameters determined in the previous work \cite{Ivanov:1997yg}.
\par
\begin{figure}[h]
    \centering
    \includegraphics[width=0.55\textwidth]{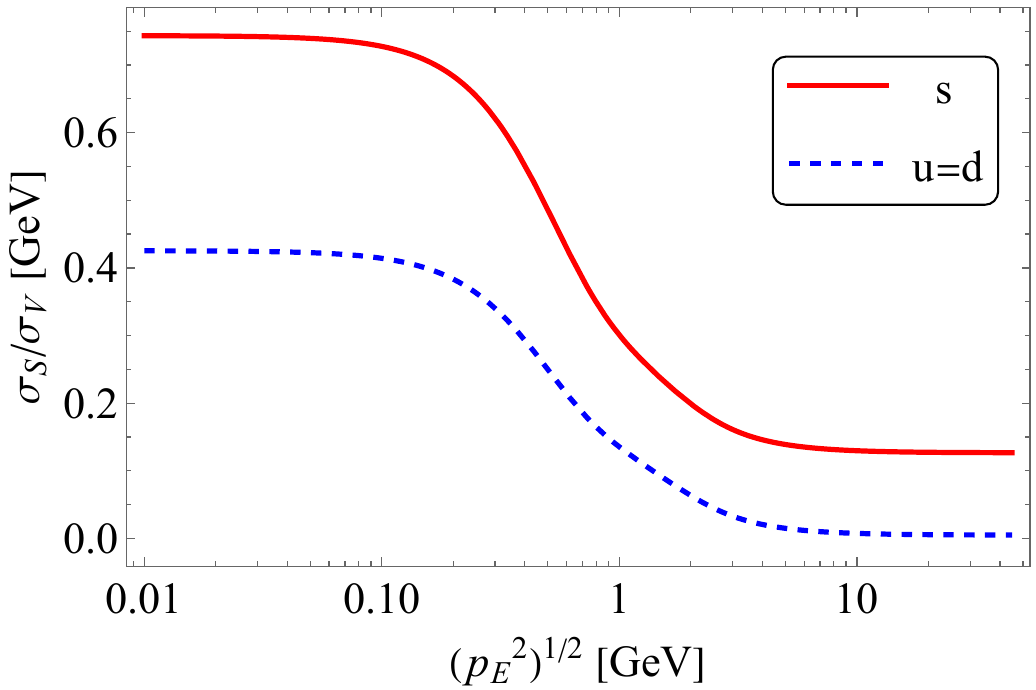} 
    \caption{Momentum dependence of $\sigma_S/\sigma_V$ in the DSE approach based on the previous work \cite{Ivanov:1997yg}.}
    \label{Fig:1}
\end{figure}

In order to illustrate typical momentum dependence of the quark propagators, $\sigma_S/\sigma_V$, corresponding to the quark mass function, is plotted in Fig.~\ref{Fig:1}, clearly showing DCSB. The parameters in Eqs.~(\ref{Eq:propV}), (\ref{Eq:propV2}), (\ref{Eq:propS}) are fixed by \cite{Ivanov:1997yg} $D=0.160~\textrm{GeV}$,
$\epsilon=10^{-4}$ and,
\bea
\tilde{m}_u=0.00897~\textrm{GeV},~b_0^u=0.131,~b_1^u=2.90,
b_2^u=0.603,~b_3^u=0.185,\label{Eq:upara}\\
\tilde{m}_s=0.224~\textrm{GeV},~b_0^s=0.105,~b_1^s=2.90,
b_2^s=0.740,~b_3^s=0.185.
\eea
We take the isospin limit for Eq.~(\ref{Eq:upara}), {\it i.e.,} the equality of the parameters for $u$ and $d$ quarks.
%============
\subsection{Dispersive part of $\bar{D}^0\to D^0$ transition}
%============
By implementing the Fierz transform, the $\Delta C=1$ operator in Eq.~(\ref{Eq:O1ope}) is rewritten by,
\bea
O_1&=&(\bar{c}^\alpha \gamma^\mu P_Lu^\alpha)(\bar{q}^\beta_j \gamma_\mu P_Lq^\beta_i).\label{Eq:O1opeprime}
\eea
Hereafter, the operator basis in Eqs.~(\ref{Eq:O1opeprime}) and (\ref{Eq:O2ope}) are used for analyzing the $D^0-\bar{D}^0$ mixing. In this case, the color flow and the flavor flow are identical in a given diagram.
\par
Contributions to the dispersive Hamiltonian are divided into four parts from the $\Delta C=1$ operators (sum over $i, j= d, s$ is not taken),
\bea
M_{12}^{ij}&=&\frac{2iG_F^2}{M_D}\lambda_i\lambda_j
\int d^4x\:\textrm{tr}[X_{\mu\nu}^{ij}(x)]
\nn\\
&\times&
\{N_cC_1^2  \bra{D^0}T[\bar{c}^\alpha(x) \gamma^\mu P_L u^\alpha(x)]
[\bar{c}^\beta(0) \gamma^\nu P_L u^\beta(0)]\ket{\bar{D}^0}
\nn\\
&+&C_2^2\bra{D^0}T
[\bar{c}^\alpha(x) \gamma^\mu P_L u^\beta(x)]
[\bar{c}^\beta(0) \gamma^\nu P_Lu^\alpha(0)]\ket{\bar{D}^0}\nn\\
&+&C_1C_2\bra{D^0}T[\bar{c}^\alpha(x) \gamma^\mu P_L u^\alpha(x)]
[\bar{c}^\beta(0) \gamma^\nu P_L u^\beta(0)]\ket{\bar{D}^0}\nn\\
&+&C_2C_1\bra{D^0}T[\bar{c}^\alpha(x) \gamma^\mu P_L u^\alpha(x)]
[\bar{c}^\beta(0) \gamma^\nu P_L u^\beta(0)]\ket{\bar{D}^0}\}.
\label{Eq:M12begin}
\eea
As shown above, the terms proportional to $C_1C_2$ and $C_2C_1$ yield identical contributions. In Eq.~(\ref{Eq:M12begin}), the intermediate flavor dependence, represented by the indices of $i$ and $j$, is included only in $\lambda_i\lambda_j$ and $X_{\mu\nu}^{ij}$ defined by,
\bea
X_{\mu\nu}^{ij}(x)=S_i(0, x)\gamma_\mu P_L S_j(x, 0)\gamma_\nu P_L.
\eea
In Eq.~(\ref{Eq:M12begin}), the trace denoted as $\textrm{tr}[X_{\mu\nu}^{ij}(x)]$ runs over only the Dirac indices, not for color or flavor indices. It should be noted that $\sigma_S$ in Eq.~(\ref{Eq:PropS}) does not contribute to $\textrm{tr}[X_{\mu\nu}^{ij}(x)]$ due to its chirality structure, similarly to the conventional OPE analysis at LO.
\par
In order to estimate Eq.~(\ref{Eq:M12begin}), one inserts vacua in between the two currents for the matrix element,
corresponding to the vacuum-insertion approximation (VIA). It is known \cite{Manohar:1998xv} that this approximation works in the large $N_c$ limit since gluon exchanges in between currents are suppressed in that limit while radiative corrections within one current are included via decay constant. Since this procedure does not affect the primary SU(3) breaking, which we aim to evaluate in the present work, VIA can capture the order of magnitude of the observable as we shall see later. When inserting vacua, care must be taken for multiple ways of contracting fermion fields in the matrix element to valence quarks in the $D$-meson states, leading to
\bea
&&\bra{D^0} T[\bar{c}^\alpha(x)
\gamma^\mu P_L
u^\alpha(x)]
[\bar{c}^\beta(0)
\gamma^\nu P_L
u^\beta(0)]\ket{\bar{D}^0}\nn\\
&\simeq&
\bra{D^0} \bar{c}^\alpha(x)
\gamma^\mu P_L
u^\alpha(x)\ket{0}
\bra{0}\bar{c}^\beta(0)
\gamma^\nu P_L
u^\beta(0)
\ket{\bar{D}^0}
\nn\\
&+&
\bra{D^0} \bar{c}^\beta(0)
\gamma^\nu P_L
u^\beta(0)\ket{0}
\bra{0}\bar{c}^\alpha(x)
\gamma^\mu P_L
u^\alpha(x)
\ket{\bar{D}^0}\nn\\
&-&
\bra{D^0} \bar{c}_a^\alpha(x)u_d^\beta(0)\ket{0}
\bra{0}\bar{c}_c^\beta(0) u_b^\alpha(x)
\ket{\bar{D}^0}
(\gamma^\mu P_L)_{ab}(\gamma^\nu P_L)_{cd}
\nn\\
&-&
\bra{D^0} \bar{c}_c^\beta(0) u^\alpha_b(x)\ket{0}
\bra{0}\bar{c}_a^\alpha(x)  u^\beta_d(0)
\ket{\bar{D}^0}
(\gamma^\mu P_L)_{ab}(\gamma^\nu P_L)_{cd},
\label{Eq:VIAstart}
\eea
where $a, b, c,$ and $d$ represent the Dirac indices. For the third and fourth terms in Eq.~(\ref{Eq:VIAstart}), one can rearrange the color indices,
\bea
[\bar{c}^\alpha u^\beta]
[\bar{c}^\beta u^\alpha]&=&
\frac{1}{N_c}[\bar{c}^\alpha u^\alpha]
[\bar{c}^\beta u^\beta]
+2[\bar{c}^\alpha  T^a_{\alpha\beta} u^\beta][\bar{c}^\gamma T^a_{\gamma\delta} u^\delta],\label{Eq:Coloriden}
\eea
which follows from $T^a_{\alpha\beta}T^a_{\gamma \delta}=(\delta_{\alpha\delta}\delta_{\gamma\beta}- \delta_{\alpha\beta}\delta_{\gamma \delta}/N_c)/2$. Since the Bethe-Salpeter amplitude is diagonal in color indices, $\bra{D^0}u^\alpha \bar{c}^\beta\ket{0}\propto \delta^{\alpha\beta}$ (see, {\it e.g.}, Ref.~\cite{Kugo:1993rd}), the matrix element of the second term in Eq.~(\ref{Eq:Coloriden}) vanishes due to $\textrm{tr}(T^a)=0$.
Furthermore, for the third and fourth terms in Eq.~(\ref{Eq:VIAstart}), one can implement the Fierz transform,
\bea
(\gamma^\mu P_L)_{ab}(\gamma^\nu P_L)_{cd}
=\frac{1}{2}
(\gamma^\mu\gamma^\rho\gamma^\nu P_L)_{ad}
(\gamma_\rho P_L)_{cb}.
\eea
By noting that parity must be preserved so that 
the matrix element of the form, $\bra{D^0} \bar{c}\gamma^\mu u\ket{0}$, vanishes, we rewrite
Eq.~(\ref{Eq:VIAstart}),
\bea
&&\bra{D^0} T[\bar{c}^\alpha(x)
\gamma^\mu P_L
u^\alpha(x)]
[\bar{c}^\beta(0)
\gamma^\nu P_L
u^\beta(0)]\ket{\bar{D}^0}\nn\\
&\simeq&
\frac{1}{4}\bra{D^0} \bar{c}^\alpha(x) \gamma^\mu\gamma^5 u^\alpha(x)\ket{0}
\bra{0}\bar{c}^\beta(0)\gamma^\nu\gamma^5 u^\beta(0)
\ket{\bar{D}^0}
\nn\\
&+&
\frac{1}{4}\bra{D^0}\bar{c}^\beta(0)\gamma^\nu\gamma^5 u^\beta(0)\ket{0}
\bra{0}\bar{c}^\alpha(x) \gamma^\mu\gamma^5 u^\alpha(x)
\ket{\bar{D}^0}\nn\\
&-&\frac{1}{2N_c}
\bra{D^0} \bar{c}^\alpha(x)
\gamma^\mu \gamma^\rho \gamma^\nu P_L
u^\alpha(0)\ket{0}
\bra{0}\bar{c}^\beta(0)
\gamma_\rho P_L
u^\beta(x)
\ket{\bar{D}^0} 
\nn\\
&-&\frac{1}{2N_c}
\bra{D^0} \bar{c}^\alpha(0) 
\gamma_\rho P_L
u^\alpha(x)\ket{0}
\bra{0}\bar{c}^\beta(x)  
\gamma^\mu \gamma^\rho \gamma^\nu P_L
u^\beta(0)
\ket{\bar{D}^0}.
\label{Eq:formula2}
\eea
Formulas similar to Eq.~(\ref{Eq:formula2}) are found for the terms proportional to $C^2_2, C_1C_2$, and $C_2C_1$ in Eq.~(\ref{Eq:M12begin}) due to the analogous procedure.
\par
\begin{figure}[t]
    \centering
    \includegraphics[width=0.85\textwidth]{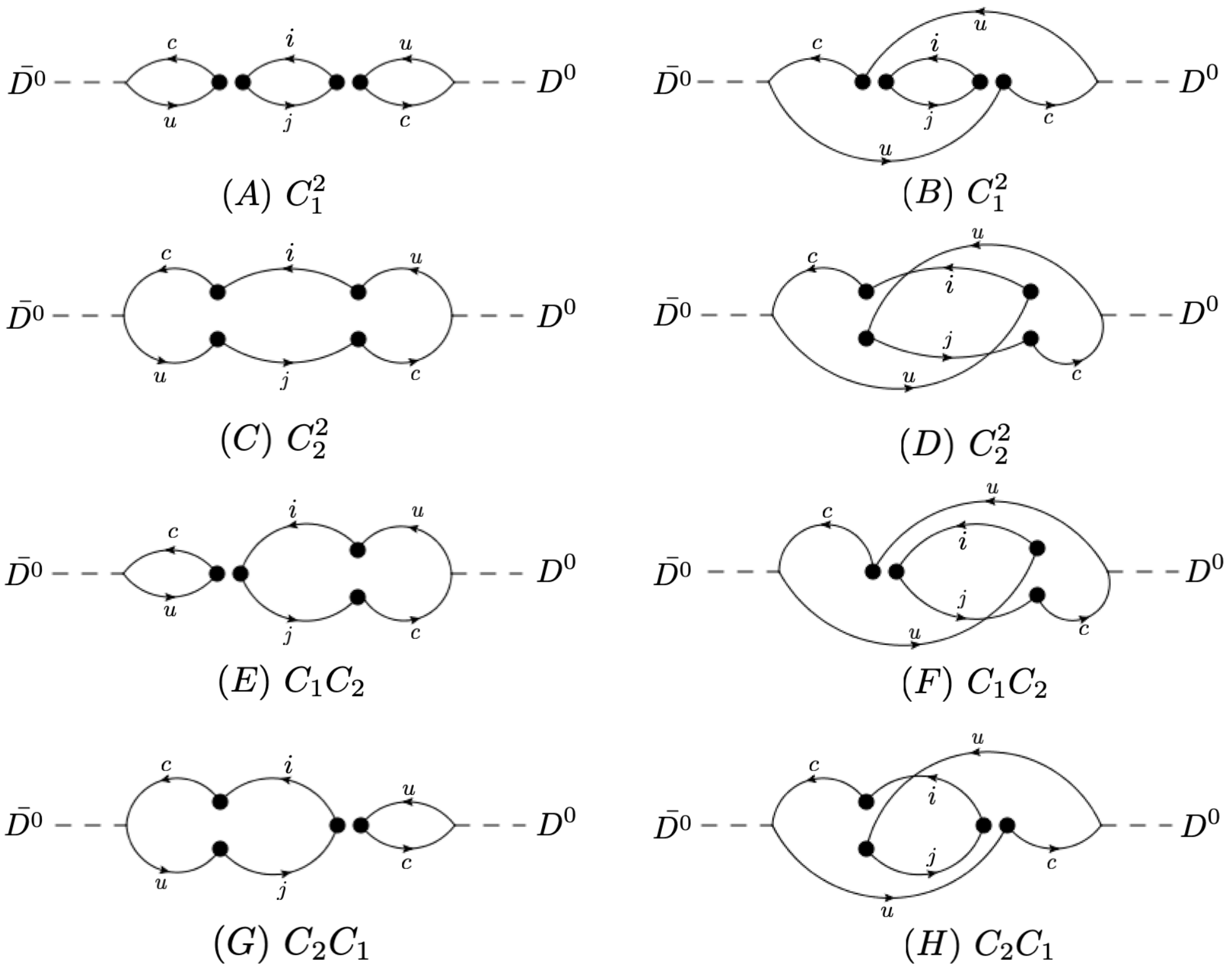} 
\caption{Diagrams for $\bar{D}^0\to D^0$ transition in the DSE approach with $i, j= d, s$. The black dots denote the Dirac matrix of the form, $\gamma^\mu P_L$. The fermion lines represent dressed quark propagators with
the arrows denoting flavor and color flows. The similar diagrams where the roles of the left and right vertices are interchanged are not displayed.}
\label{Fig:2}
\end{figure}

By using the decomposition in Eq.~(\ref{Eq:formula2}),
one finds that the Hamiltonian in Eq.~(\ref{Eq:effHdef}) includes diagrams in Fig.~\ref{Fig:2}. As shown in the figure, there are eight types of diagrams, corresponding to terms proportional to $C_1^2, C_2^2, C_1C_2$, and $C_2C_1$, each of which has two types of different contractions, explicitly displayed in Fig.~\ref{Fig:2}. $N_c$ dependence of each diagram can be read off by counting the number of color traces. In addition to those diagrams, there are further eight diagrams, where the roles of the two $\Delta C=1$ vertices are simply interchanged for each of A-H, which we denote A$^\prime$-H$^\prime$.
%==========
\par
By evaluating the left-side of the diagrams plus corresponding primed ones in Fig.~\ref{Fig:2}, partial contributions to $M_{12}$ are,
\bea
M_{12}^{ij}|_{A+A^\prime}&=&
N_cC_1^2 N_{I}^{ij},\quad
M_{12}^{ij}|_{C+C^\prime}=
\frac{C_1^2}{N_c}N_{I}^{ij},\\
M_{12}^{ij}|_{E+E^\prime}&=&
C_1C_2N_{I}^{ij},\quad
M_{12}^{ij}|_{G+G^\prime}=C_2C_1N_{I}^{ij}.
\eea
where we introduced,
\bea
N_I^{ij}&=&
\frac{iG_F^2}{2M_D}\lambda_i\lambda_j
\int d^4x\:\textrm{tr}[X_{\mu\nu}^{ij}(x)]
[\bra{D^0}\bar{c}^\alpha(x) \gamma^\mu \gamma^5 u^\alpha(x)\ket{0}
\nn\\
&&\times \bra{0}\bar{c}^\beta(0) \gamma^\nu \gamma^5 u^\beta(0)\ket{\bar{D}^0}+(x\leftrightarrow 0)].\label{Eq:NIijeq}
\eea
In the above relation, the term denoted as $(x\leftrightarrow 0)$ represents a primed diagram. The matrix elements in Eq.~(\ref{Eq:NIijeq}) are written by the decay constant,
\bea
\bra{D^0}\bar{c}^\alpha(x) \gamma^\mu \gamma^5 u^\alpha(x)\ket{0}\bra{0}
\bar{c}^\beta(0) \gamma^\nu \gamma^5 u^\beta(0)\ket{\bar{D}^0}=f_{D^0}^2
p^\mu p^\nu e^{ip\cdot x}.\label{Eq:mateleimp}
\eea
In Eq.~(\ref{Eq:mateleimp}), $p$ is the momentum of the $D$ mesons satisfying the on-shell condition, $p^2=M_{D^0}^2$.
\par
One can similarly evaluate the right-side of the diagrams plus corresponding primed ones in Fig.~\ref{Fig:2}. The third and fourth terms in Eq.~(\ref{Eq:formula2}) are simplified via the relation for gamma matrices ($\epsilon^{0123}=-\epsilon_{0123}=1$),
\bea
\gamma^\mu \gamma^\rho \gamma^\nu =g^{\mu\rho}\gamma^\nu+g^{\rho\nu}\gamma^\mu-g^{\mu\nu}\gamma^\rho-i\epsilon^{\mu\nu\rho\sigma}\gamma_\sigma\gamma_5.
\eea
Using the mentioned algebra, we obtain,
\bea
M_{12}^{ij}|_{B+B^\prime}&=&C_1^2N_{J}^{ij},\quad
\hspace{5.3mm}
M_{12}^{ij}|_{D+D^\prime}=C_2^2N_{J}^{ij},\\
M_{12}^{ij}|_{F+F^\prime}&=&\frac{C_1C_2}{N_c}N_{J}^{ij},\quad
M_{12}^{ij}|_{H+H^\prime}=\frac{C_2C_1}{N_c}N_{J}^{ij}.
\eea
In the above relations, the following quantity is introduced,
\bea
N_J^{ij}&=&-i\frac{G_F^2}{4M_D}\lambda_i\lambda_j
\int d^4x\:\textrm{tr}[X_{\mu\nu}^{ij}(x)]
(g^{\mu\rho}g^{\nu\sigma}+g^{\nu\rho}g^{\mu\sigma}
-g^{\nu\mu}g^{\rho\sigma})\nn\\
&\times&
[\bra{D^0}\bar{c}^\alpha(x)\gamma_\rho\gamma_5 u^\alpha(0)\ket{0}
\bra{0}\bar{c}^\beta(0)\gamma_\sigma\gamma_5 u^\beta(x)\ket{\bar{D}^0}+(x\leftrightarrow 0)].\label{Eq:Mateleeq}
\eea
The matrix elements in Eq.~(\ref{Eq:Mateleeq}) is evaluated in the approximation where light quark field is local, $u(x)\simeq u(0)$. This procedure leads to a relation similar to Eq.~(\ref{Eq:mateleimp}),
\bea
\bra{D^0}\bar{c}^\alpha(x)\gamma_\rho\gamma_5 u^\alpha(0)\ket{0}
\bra{0}\bar{c}^\beta(0)\gamma_\sigma\gamma_5 u^\beta(x)\ket{\bar{D}^0}\simeq f_{D^0}^2p_\rho p_\sigma
e^{ip\cdot x}.
\eea
The contribution of A$^\prime$ to $M_{12}$ is merely equal to the one from A, {\it i.e.,} $M_{12}^{ij}|_{A}=M_{12}^{ij}|_{A^\prime}$. The analogous relations are valid also for B, C, $\cdots$, H and the corresponding primed ones.
\subsection{Expressions for $M_{12}$}
By assembling the obtained results, the sum over the diagrams and intermediate flavors is carried out straightforwardly,
\bea
M_{12}^{(\alpha)}&\equiv &M_{12}^{A+A^\prime+C+C^\prime+E+E^\prime+G+G^\prime}
=2G_F^2 f_{D^0}^2M_{D^0}^3\lambda_s^{2}\left(N_cC_1^2+\frac{C_2^2}{N_c}+2C_1C_2\right)I,\label{Eq:M12sum1}\\
M_{12}^{(\beta)}&\equiv&
M_{12}^{B+B^\prime+D+D^\prime+F+F^\prime+H+H^\prime}
=4G_F^2 f_{D^0}^2M_{D^0}^3\lambda_s^{2}
\left(C_1^2+C_2^2+\frac{2}{N_c}C_1C_2\right)J,\label{Eq:M12sum2}
\eea
where the unitarity of the CKM matrix is used in a way similar to the OPE analysis. In Eqs.~(\ref{Eq:M12sum1}), (\ref{Eq:M12sum2}),
$I$ and $J$ are momentum integrals that consist of contributions which depend on intermediate flavors,
\bea
K=K_{dd}+K_{ss}-K_{sd}-K_{ds}.\quad (K=I, J)\label{Eq:Kint}
\eea
It should be noted that $I$ and $J$ work as SU(3) breaking factors corresponding to $[(m_s^2-m_d^2)/m_c^2]^2$ in Eq.~(\ref{Eq:M12SU3}) for the perturbative case, up to normalization. Each term in Eq.~(\ref{Eq:Kint}) is defined by,
\bea
I_{ij}&=&i\int\frac{d^4l}{(2\pi)^4}
\left[\frac{l^2+p\cdot l}{M_{D^0}^2}
-2\frac{(p\cdot l)^2}{M_{D^0}^4}\right]
\sigma_{V}^i(-(l-p)^2)
\sigma_{V}^j(-l^2),\label{Eq:Ires}\\
J_{ij}&=&i\int\frac{d^4l}{(2\pi)^4}\frac{p\cdot l}{M_{D^0}^2}
\left(\frac{p\cdot l-M_{D^0}^2}{M_{D^0}^2}\right)\sigma_{V}^i(-(l-p)^2)
\sigma_{V}^j(-l^2).\label{Eq:Jres}
\eea
We evaluate the integrals in Eqs.~(\ref{Eq:Ires}), (\ref{Eq:Jres}) at the rest frame of the $D$ meson, $p=(M_{D^0}, \mathbf{0})$, by performing the Wick rotation, where relevant Euclidean variables are introduced by, $l_0=il_4$ and $l^2=-l^2_E$ (see, {\it e.g.}, Ref.~\cite{Roberts:2000aa}). The expressions read,
\bea
I_{ij}&=&\int_0^{\Lambda_{\rm cut}}da
\int_0^\pi d\omega
\frac{a^3}{4\pi^3}\sin^2\omega
\left(
\frac{a^2}{M_{D^0}^2}
-\frac{2a^2}{M_{D^0}^2}\cos^2\omega
-i\frac{a}{M_{D^0}}\cos\omega
\right)\nn\\
&\times&\sigma_V^i(a^2-M_{D^0}^2+2iM_{D^0}a\cos\omega)\sigma_V^j(a^2),\label{Eq:intIform}\\
J_{ij}&=&\int_0^{\Lambda_{\rm cut}}da
\int_0^\pi d\omega
\frac{a^3}{4\pi^3}\sin^2\omega
\frac{a\cos\omega}{M_{D^0}}\left(
\frac{a\cos\omega}{M_{D^0}}+i
\right)\nn\\
&\times&
\sigma_V^i(a^2-M_{D^0}^2+2iM_{D^0}a\cos\omega)\sigma_V^j(a^2),
\label{Eq:intJform}
\eea
regulated by ultraviolet cutoff ($\Lambda_{\rm cut}$). It should be noted that $I$ and $J$ are finite and do not depend on the cutoff for sufficiently large $\Lambda_{\rm cut}$ since contributions from large momenta cancel in Eq.~(\ref{Eq:Kint}). Equations (\ref{Eq:intIform}), (\ref{Eq:intJform}) can be evaluated with the vector part of the quark propagators in Eqs.~(\ref{Eq:propV}), (\ref{Eq:propV2}). The expression for the entire $M_{12}$ is obtained by summing over Eqs.~(\ref{Eq:M12sum1}), (\ref{Eq:M12sum2}), {\it i.e., }$M_{12}=M_{12}^{(\alpha)}+M_{12}^{(\beta)}$.
%============
\subsection{Numerical result}
%============
For numerical input, we consider the following parameter set:
\begin{itemize}
\item[(i)] $f_{D^0}=0.212~\textrm{GeV}$ (lattice QCD \cite{Carrasco:2014poa})
and $M_{D^0}=1.86484~\textrm{GeV}$ (PDG \cite{ParticleDataGroup:2024cfk})
\item[(ii)] $f_{D^0}=0.227~\textrm{GeV}$
and $M_{D^0}=1.87~\textrm{GeV}$ (DSE approach \cite{Ivanov:1997yg})
\item[(iii)] $f_{D^0}=0.272~\textrm{GeV}$ 
and $M_{D^0}=1.87~\textrm{GeV}$ (DSE approach \cite{Ivanov:1997yg})
\end{itemize}
The decay constants in the second and third cases in the DSE approach are larger than the value in lattice QCD by 7$\%$ and $28\%$, respectively. As for the ultraviolet cutoff in Eqs.~(\ref{Eq:intIform}), (\ref{Eq:intJform}), we fix $\Lambda_{\rm cut}=10^{2}~\textrm{GeV}$. We verified that the numerical result is reasonably stable under the variation of $\Lambda_{\rm cut}$. Furthermore, the following two cases are discussed,
\bea
\textrm{Case [A]}:&&\:
\textrm{VIA, local limit of $u(x)$,}~
\Delta C=1\textrm{~weak~coupling}:
~C_1(m_c), ~~C_2(m_c)
\nn\\
\textrm{Case [B]}:&&\:
\textrm{VIA, local limit of $u(x)$, no QCD for}~\Delta 
C=1\textrm{~weak~coupling}:
~C_1=0, ~~C_2=1,\nn
\eea
where specific values of $C_{1}(m_c)$ and $C_2(m_c)$ are fixed to the ones for the OPE analysis in Sec.~\ref{Sec:IIC}. Other parameters such as CKM matrix elements are also fixed as the previous values.
%================
\begin{table}[t]
%================
\centering
\caption{
$|x|$ in this work compared with previous results.
The first to the sixth rows show the results in the present work based on Cases [A] and [B] (see main texts) and varying $f_{D^0}$ and $M_{D^0}$. The seventh, eighth and ninth rows give the previous OPE-based results. For the seventh row, the result of $(\Delta m)_{HqEFT}\approx(0.9-3.5)\cdot 10^{-17}~\textrm{GeV}$ \cite{Ohl:1992sr} is converted to $|x|$ via the $D^0$-meson lifetime data in Ref.~\cite{HeavyFlavorAveragingGroupHFLAV:2024ctg}. The tenth row gives the experimental value.}
\label{Tab:II}
\begin{tabular}{crlll}
\hline\hline
This work & $|x|=$& $1.7$&$ \times 10^{-3} $  & DSE ([A] $f_{D^0}=0.212~\textrm{GeV}, M_{D^0}=1.86484~\textrm{GeV}$)
\\\hline
This work &$|x|=$ & $1.3$&$ \times 10^{-3} $ &  DSE ([B] $f_{D^0}=0.212~\textrm{GeV}, M_{D^0}=1.86484~\textrm{GeV}$)
\\\hline
This work & $|x|=$& $2.0$&$ \times 10^{-3} $  & DSE ([A] $f_{D^0}=0.227~\textrm{GeV}, M_{D^0}=1.87~\textrm{GeV}$) 
\\\hline
This work &$|x|=$ & $1.5$&$ \times 10^{-3} $ &  DSE ([B] $f_{D^0}=0.227~\textrm{GeV}, M_{D^0}=1.87~\textrm{GeV}$ )
\\\hline
This work & $|x|=$& $2.9$&$ \times 10^{-3} $  & DSE ([A] $f_{D^0}=0.272~\textrm{GeV}, M_{D^0}=1.87~\textrm{GeV}$ ) 
\\\hline
This work &$|x|=$ & $2.2$&$ \times 10^{-3} $ &  DSE ([B] $f_{D^0}=0.272~\textrm{GeV}, M_{D^0}=1.87~\textrm{GeV}$)
\\\hline
Ref.~\cite{Ohl:1992sr}&$|x|=$ & $(5.6-22) $&$\times 10^{-6}$& OPE (HQET)
\\\hline
Ref.~\cite{Golowich:2005pt} &$|x|=$ &$6$&$\times 10^{-7}$& OPE (NLO)
\\\hline
Ref.~\cite{Melic:2024oqj} &$|x^{NLC}|=$ & $7.7$&$\times 10^{-6}$& OPE (Nonlocal condensate)\\\hline
HFLAV \cite{HeavyFlavorAveragingGroupHFLAV:2024ctg}
&$x=$&$(4.07\pm 0.44)$&$\times 10^{-3}\qquad$& Experiment \\\hline\hline
\end{tabular}
\end{table}
%================
\begin{table}[t]
%================
\centering
\caption{Partial contribution to $x$ in this work separately taking the left-side ($x^{(\alpha)}$) and right-side ($x^{(\beta)}$) of the diagrams in Fig.~\ref{Fig:2}.}
\label{Tab:III}
\begin{tabular}{rlrll}
\hline\hline
$|x^{(\alpha)}|=$& $1.2\times 10^{-6}$& $|x^{(\beta)}|=$ & $1.7\times 10^{-3}$ $\quad$
& [A] $f_{D^0}=0.212~\textrm{GeV}, M_{D^0}=1.86484~\textrm{GeV}$
\\\hline
$|x^{(\alpha)}|=$& $0.20\times 10^{-3}$& $|x^{(\beta)}|=$ & $1.5\times 10^{-3}$ $\quad$
& [B] $f_{D^0}=0.212~\textrm{GeV}, M_{D^0}=1.86484~\textrm{GeV}$
\\\hline
$|x^{(\alpha)}|=$& $1.4\times 10^{-6}$& $|x^{(\beta)}|=$ & $2.0\times 10^{-3}$ $\quad$
& [A] $f_{D^0}=0.227~\textrm{GeV}, M_{D^0}=1.87~\textrm{GeV}$ 
\\\hline
 $|x^{(\alpha)}|=$& $0.24\times 10^{-3}$& $|x^{(\beta)}|=$ & $1.8\times 10^{-3}$ $\quad$
& [B] $f_{D^0}=0.227~\textrm{GeV}, M_{D^0}=1.87~\textrm{GeV}$
\\\hline
$|x^{(\alpha)}|=$& $2.1\times 10^{-6}$& $|x^{(\beta)}|=$ & $2.9\times 10^{-3}$ $\quad$
& [A] $f_{D^0}=0.272~\textrm{GeV}, M_{D^0}=1.87~\textrm{GeV}$ 
\\\hline
$|x^{(\alpha)}|=$& $0.34\times 10^{-3}$& $|x^{(\beta)}|=$ & $2.5\times 10^{-3}$ $\quad$
& [B] $f_{D^0}=0.272~\textrm{GeV}, M_{D^0}=1.87~\textrm{GeV}$
\\\hline\hline
\end{tabular}
\end{table}
\par
The numerical results in the DSE approach are exhibited and compared with the previous works in Table~\ref{Tab:II}. One finds that the results in the DSE approach are $|x|=(1.3-2.9)\times 10^{-3}$, the order of magnitude comparable to the experimental value. This leads to a certain improvement compared to the OPE-based approaches, in which case the order of magnitude is far below the data. In addition, we find that the results in the DSE approach are sensitive to the parameters of $f_{D^0}$, $C_1$ and $C_2$ and slightly to
$M_{D^0}$.
\par
Numerical results which separately consider the left-side and right-side of the diagrams in Fig.~\ref{Fig:2} are respectively extracted by,
\bea
x^{(\alpha)}=\frac{2M_{12}^{(\alpha)}}{\Gamma},\quad
x^{(\beta)}=\frac{2M_{12}^{(\beta)}}{\Gamma},
\eea
where $M_{12}^{(\alpha)}$ and $M_{12}^{(\beta)}$ are defined in Eqs.~(\ref{Eq:M12sum1}), (\ref{Eq:M12sum2}). The results of $x^{(\alpha)}$ and $x^{(\beta)}$ are exhibited in Table~\ref{Tab:III} with different input parameters for $f_{D^0}, M_{D^0}, C_1$ and $C_2$, as well as Table~\ref{Tab:II}. We find that for Case [A], the value of $|x^{(\alpha)}|$ is of $\mathcal{O}(10^{-6})$, much smaller than $|x^{(\beta)}|=\mathcal{O}(10^{-3})$. This is caused by the accidental cancellation for the Wilson coefficients in Eq.~(\ref{Eq:M12sum1}),
\bea
N_c C_1^2+\frac{C_2^2}{N_c}+2C_1C_2\approx 0.002\ll 
2\left(C_1^2+C_2^2+\frac{2}{N_c}C_1C_2\right)\approx
2.3.\label{Eq:Wilsoncomp}
\eea
As for Case [B] in Table~\ref{Tab:III}, one finds that $|x^{(\alpha)}|=\mathcal{O}(10^{-4})$ and $|x^{(\beta)}|=\mathcal{O}(10^{-3})$. The slightly small value of $|x^{(\alpha)}|$ can be partially understood by taking $C_2\to 1$ and $C_1\to 0$ in Eq.~(\ref{Eq:Wilsoncomp}), leading to $N_c^{-1}<2$.
%==================
\section{Conclusion}\label{Sec:IV}
%==================
In this work, we have discussed the $D^0-\bar{D}^0$ mixing in such a way that chiral symmetry breaking is properly taken into account. In the conventional OPE approach, the factor proportional to SU(3) breaking, which originates from the GIM mechanism, gives the extreme suppression so that $x$ at the lowest order is significantly smaller than the HFLAV data. We have assessed how this result is altered when the current-quark masses are replaced by the constituent-quark masses. It is found that the order of the magnitude is enhanced typically by $\mathcal{O}(10^{2})$ although the detailed values are sensitive to the constituent-quark masses.
\par
In order to incorporate the entire momentum dependence of the quark propagators, not just replacing the quark masses, the DSE approach is discussed to evaluate the mixing parameter. On the basis of the parametrization for quark propagators in the previous works \cite{Burden:1991gd, Burden:1995ve, Kalinovsky:1996ii, Ivanov:1997yg, Ivanov:1998ms}, DCSB from primary SU(3) breaking is appropriately taken into account. With the $\Delta C=1$ local operators, the $\bar{D}^0\to D^0$ transition is evaluated in the VIA, formulated by the decay constant of $D^0$ meson, as well as relevant momentum integrals. 
\par
In the local limit of the fermion field corresponding to the valence light quark of $D$ meson, the dimensionless mass-difference observable is numerically investigated. It is shown that the result is sensitive to the values of the decay constant, $\Delta C=1$ Wilson coefficients and (slightly) $D^0$-meson mass. The contributions of the two different types of diagrams, denoted as $x^{(\alpha)}$ and $x^{(\beta)}$, are discussed separately. We have found that for $C_1=C_1(m_c)$ and $C_2=C_2(m_c)$, the accidental cancellation occurs between some of the diagrams, resulting in suppression for $x^{(\alpha)}$. For $C_1=0$ and $C_2=1$, $|x^{(\alpha)}|< |x^{(\beta)}|$ is still realized, partially understandable via
the large-$N_c$ counting. 
\par
The obtained numerical results for the observable read $|x|=(1.3-2.9)\times 10^{-3}$, where the uncertainty range comes from the variation of $f_{D^0}$, $M_{D^0}, C_1$ and $C_2$, while experimental 95$\%$ confidence-level interval is given by $x=[3.20, 4.93]\times 10^{-3}$ \cite{HeavyFlavorAveragingGroupHFLAV:2024ctg}.
As such, the order of magnitude of $x$ is comparable to the data in the DSE approach, which leads to a certain improvement from the OPE-based approaches. Further scrutinization of the result is foreseen \cite{PREP} in such a way that takes account of nonlocality for the light quark field by means of the Bethe-Salpeter approach.
%==================
\section*{Acknowledgments}
%==================
The authors would like to thank Shinya Matsuzaki for reading the manuscript and giving us useful comments. This work is supported by the National Science Foundation of China (NSFC) under Grant No.~12405111 and the Seeds Funding of Jilin University.
%==================
\section*{Data availability}
%==================
The data supporting this study's findings are available within the article.
%=============
\appendix
%=============
\section{RGE AT LO IN OPE}\label{App:A}
Here we denote $\mu_L=3~\mathrm{GeV}$, $\mu_c=m_c(m_c)$ and $\mu_b=m_b(m_c)$, with the masses given by the $\overline{\rm MS}$ scheme at the scale of the charm-quark mass. The bag parameters from the ETM collaboration \cite{Carrasco:2015pra} are,
\bea
B_1(3~\mathrm{GeV})&=&0.757(27)(4),\label{Eq:Bag1}\\
B_2(3~\mathrm{GeV})&=&0.65(3)(2),\label{Eq:Bag2}\\
B_3(3~\mathrm{GeV})&=&0.96(8)(2).\label{Eq:Bag3}
\eea
In order to evaluate Eq.~(\ref{Eq:M12SU3}), $B_1(\mu_c)$ and $B_2(\mu_c)$ are required. These quantities can be straightforwardly obtainable from the discussion in Ref.~\cite{Buras:2001ra}, resulting in at LO,
\bea
B_1(\mu_c)&=&
B_1(\mu_{\rm L})\left[\frac{\alpha_s(\mu_c)}{\alpha_s(\mu_{\rm L})}\right]^{\frac{\gamma^{\rm VLL}}{2\beta_0(4)}},\nn\\
B_2(\mu_c)&=&
(U_{11}^{\rm SLL}+4U_{21}^{\rm SLL})
B_2(\mu_{\rm L})
-\frac{8}{5}U_{21}^{\rm SLL}B_3(\mu_{\rm L}),
\eea
where the factors in regards to the RGE are \cite{Buras:2001ra},
\bea
\beta_0(N_f)&=&11-\frac{2}{3}N_f\label{Eq:beta0},\\
\gamma^{\rm VLL}&=&4,\\
U^{\rm{SLL}}&=&V^{\rm SLL}\rm{diag}\left\{
\left[\frac{\alpha_s(\mu_{\rm_c})}{\alpha_s(\mu_{\rm L})}\right]^{\frac{{\gamma^{SLL}}}{2\beta_0(4)}}\right\}(V^{\rm SLL})^{-1},\\
\vec{\gamma}^{\mathrm{SLL}}&=&\left(\frac{2}{3}+\frac{2}{3}\sqrt{241},\quad \frac{2}{3}-\frac{2}{3}\sqrt{241} \right),\\
V^{\mathrm{SLL}}&=&
\begin{pmatrix}
-60 &-60\\
16+\sqrt{241} & 16-\sqrt{241}
\end{pmatrix}.
\eea
In Eq.~(\ref{Eq:beta0}), $N_f$ represents the number of quarks.\par
As for the RGE for the Wilson coefficients of the current-current operators, these are obtained in a standard way in Ref.~\cite{Buchalla:1995vs} and references therein: By introducing,
\bea
C_\pm(\mu) = C_2(\mu)\pm C_1(\mu),\label{Eq:Cpm}
\eea
and using $C_1(M_W)=0$ and $C_2(M_W)=1$ at LO,
\bea
C_{\pm}(\mu_c)&=&
\left[\frac{\alpha_s(\mu_b)}{\alpha_s(\mu_c)}\right]
^{\frac{\gamma_{\pm}}{2\beta_0(4)}}
\left[\frac{\alpha_s(M_W)}{\alpha_s(\mu_b)}\right]
^{\frac{\gamma_{\pm}}{2\beta_0(5)}},\label{Eq:WilLO}\\
\gamma_{\pm}&=&\pm 6\frac{N_c\mp 1}{N_c}.
\eea
The values of $C_1(\mu_c)$ and $C_2(\mu_c)$ are determined by solving Eq.~(\ref{Eq:Cpm}) with Eq.~(\ref{Eq:WilLO}). Numerically, those LO values are $C_1(\mu_c)\approx -0.35$ and $C_2(\mu_c)\approx1.13$.
%%%%%%%%%%%%
%%% References %%%
%%%%%%%%%%%%

\end{document}